# Spin Based Biosensor with Hard Axis Assist for Enhanced Sensitivity

B. Manna[1], A. K. Muhopadhyay[1] and M. Sharad[1]

[1] *Department of Electronics& Electrical Communication Engineering , IIT Kharagpur, West Bengal, India*

*Abstract*— We demonstrate the influence of hard axis assist on an in-plane polarized nanomagnet layer to greatly enhance the sensitivity of a magnetic nano particle (MNP) based MTJ biosensor. The hard axis assist has been provided to the sensing layer in the forms of spin Hall Effect (SHE) induced spin injected torque and stress based effective magnetic field induced torque separately. Present work mainly focuses on the efficient and qualitative detection of a single magnetic bead with the aim of detecting a single bimolecular recognition event at an extremely low analyte concentration. Interfacial spin current arising from spin orbit metal is expected to impose a torque on the free layer along sensitive direction improving the signal strength by a factor of ~ 6.5 for a 100 nm bead at a height of 500 nm above the sensor surface. Furthermore, the potentiality of a multiferroic composite consisting of a piezoelectric layer coupled with magnetostrictive CoFeB based MTJ has been investigated in the Biosensing applications. An external stress voltage of 500 mV has been observed to be sufficient to enhance the sensitivity (~ 6 times). The use of nanoscaled spin devices and the absence of external magnetic field operated magnetization rotation facilitates in achieving highly compact and extremely low power designs. This establishes the possibility of utilizing the present schemes for advanced, highly sensitive, miniaturized and low power bioassaying system-on-chip applications. Numerical results were compared with some earlier reported experimental results to validate our proposed model.

## I. INTRODUCTION

One of the key Challenges in the present day medical diagnostic is the rapid, quantitative and accurate detection of the disease biomarkers or pathogens in the biological samples. Different types of labels such as fluorophore, quantum dot, magnetic particle etc. are used to tag the biomarkers for their efficient detection in state-of-the-art ELISA techniques. Although, fluorescence based bioassaying methods received wide acceptance, these schemes suffer from relatively higher detection threshold (typically $10^4$ bio-molecules needed for obtaining useful SNR at output) due to the involvement of crosstalk and bleaching in conventional optical detection systems [Schena 2000]. The advancement in magneto-nanotechnology has made it possible to improve this lower limit of the bio-detection significantly by utilizing MNPs as labels due to their dimensional alikeness with disease biomarkers and least chance of a biological sample having magnetic contamination in magnetic immunoassay. All of these magnetic assaying schemes basically rely on the measurement of magnetic stray field emanated from specifically bound MNPs to the biologically active sensor surface. Enormous research progress in the design and development of highly sensitive affinity based magneto resistive sensors for clinical diagnostics and bioassaying applications has been witnessed in the recent times [Baselt 1998, Rife 2003, Graham 2004, Wang 2014]. Basic sensing mechanism involves detection of biorecognition events between probes patterned on the sensor surface and tagged target biomolecules near the surface. Although, Spin valves (SVs) with superparamagnetic bead as labels have been widely used for the detection of bioagents, they provide a very small signal strength for smaller label moments because of their small MR ratio (~10%) and requires a subsequent complex circuit arrangement for signal amplification leading to high power consumption [Ferreira 2003, Janssen 2008]. This problem can be alleviated by replacing SVs with another emerging spin device called MTJ as the basic sensing element owing to its huge TMR ratio (more than 300% with proper selection of MgO crystal orientation), small size and lab on chip compatibility [Albon 2009, Shen 2008, Lee 2007].

Various techniques and magnetic field arrangements have been adopted to enhance the sensitivity of the MTJ based biosensors [Lian 2012, Wang 2005, Megens 2005]. However, SHE induced torque driven in-plane hard axis (y direction) assist directly on the free layer has not been explored yet. Recently, electric current induced spin orbit torque has been considered as one of the most promising ways to control the magnetization dynamics of a ferromagnetic layer adjacent to the spin hall layer [Liu 2012]. Spin orbital interaction at the heavy metal-ferromagnet stack solely decides the orientation of the free layer magnetization direction via two different attributes: SHE and Rashba effect. SHE is responsible for displacing magnetization along in-plane hard axis direction for a charge current flowing through magnet's easy axis (z), whereas, latter one though weak, attempts to make it out of plane. Apart from the SHM based scheme, there exists another possibility of increasing the sensitivity of the magnetic biosensor using emerging strain based approach. Successful fabrication and testing of magnetostrictive CoFeB layer based MTJ structure has made the concept feasible in the sensor design along with their use in various storage and switching applications [Wang 2005]. A thin piezoelectric layer (PZT material in this case) elastically coupled with the CoFeB when subjected to a tiny voltage of few milivolts can transfer the generated stress to its adjacent magnetostrictive film [Roy 2015]. The resultant uniaxial stress subsequently exerts an effective torque on the free CoFeB layer of the MTJ along the in-plane hard axis direction causing fluctuation of the sensing layer magnetization orientation. Absence of the external magnetic field to govern the magnetization dynamics in our proposed schemes ensures great compactness and energy efficient design that are necessary parameters for lab on chip applications.

Present work emphasizes on a simple analytical model based design and simulation of an elliptical shaped MTJ biosensor subjected to SHE and strain induced hard axis assist separately. Two proposed devices implementing these novel ideas in biosensor applications has been demonstrated in the section II. Section III focuses on the result and discussion part and section IV finally concludes on this work.

## II. PROPOSED MODEL

### A. Design and Operation

A typical TMR biosensor with a single super paramagnetic nano bead immobilized on the free layer surface is shown in fig.1 (a). Here we consider an MTJ with two constituent ferromagnetic layers separated by a thin MgO barrier as a basic sensor device. The principle of such a biosensing scheme for the detection of antigen-antibody capturing event utilizing biotin-streptavidin chemistry on the biologically activated gold coated surface has been depicted in this figure [xxx]. Dimension of all three layers along major and minor axes are l and w respectively as shown in fig.1 (b). Orientation of pinned layer magnetization $M_P$ is exchanged biased to the y direction by means of some strong exchange coupling. Let $\theta_F$ and $\theta_P$ be the respective angles subtended by free and pinned layer magnetizations with +z axis at any instant of the time. Huge shape and crystalline anisotropy of the soft ferromagnetic material aligns the free layer magnetization $M_F$ along its major axis.

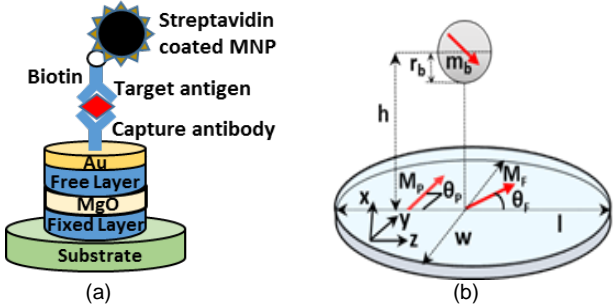

Fig. 1. (a) Layered MTJ Biosensor platform for magnetic nanobead detection. (b) Sensor strip showing dimensions and magnetization orientations of different nanomagnets. Free layer center is assumed to be at the origin.

The resulting sensor response for the two ferromagnetic layers with such an orthogonal orientation is shown in figure 2 (a). The proposed analytical model has been calibrated with the experimental data using same parameters as reported earlier [Albon 2009]. Perpendicular arrangement of the pinned and free layer magnetizations has been seen to produce a linear change in sensitivity for up to a few hundred Oersteds of applied magnetic field. This highly linearized segment of the transfer curve can be utilized for the MNP detection.

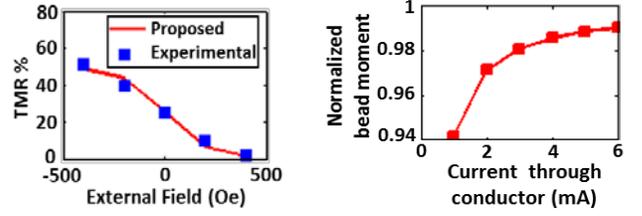

Fig. 2. (a) Calibration of the analytical model with experimental data [Albon 2009]. (b) Variation of normalized bead magnetic moment for different values of conductor current.

In reality, output of the sensor strongly rely on the combined dipole fields produced by the ensemble of immobilized MNPs. However, for the purpose of quantitative detection, analysis with a single magnetic bead has been carried out extensively. Magnetic moment, $m_b$ of an isolated, randomly oriented spherical superparamagnetic particle can be modeled by using the Langevin function [Wang 2008]. Nano bead is magnetized mainly by a current carrying conductor generating a magnetic field at the rate of ~ 3.5 Oe/mA in x-y plane normal to the sensor surface. However, the influence of small dipole fields from pinned and free layers to nano bead have also been taken into the account for the purpose of making the model more accurate and comprehensive. Magnetizing field at the particle position (h, $y_P$, $z_P$) due to rectangular conductor carrying current of $I_C$ along y direction can be computed following the Bio-Savart's law [Megens 2007]-

$$H_{CZ} = -\frac{J_C}{2\pi}\left\{\left[(x-h)\arctan(\frac{z-z_P}{x-h})+(\frac{z-z_P}{2})\ln((z-z_P)^2+(x-h)^2)\right]_{z_1}^{z_2}\right\}_{x_1}^{x_2}$$

(1)

Similar result for $H_{CX}$ can be obtained just by interchanging $x_{1,2}$ and $z_{1,2}$ and dropping the negative sign in $H_{CZ}$ expression. It is observed from fig. 2 (b) that the normalized moment tends to saturate for conductor current exceeding a value of 3 mA and hence higher value of $I_C$ will cause more power dissipation without making a significant improvement in sensitivity. This magnetized bead will in turn exert a stray field on the sensor surface which will be detected by the free sensing layer in terms of its change in $\theta_F$. It is worthwhile to mention that the large demagnetizing effect in out of plane direction makes the MTJ sensor almost insensitive to the field produced by the conductor current. Hence, after successful biological bonding of nano bead on the sensor surface, total magnetic field distribution $H_{Tot}$ on the free layer gets affected by the bead induced stray field $H_{Bead}$.

$$H_{Tot} = H_{Stray-P} + H_{Bead} \quad (2)$$

Where $H_{Stray-P}$ is the stray field produced by pinned layer on the sensing layer. The difference of $H_{Tot}$, in the absence and presence of the bead will change the magnetization orientation of the free layer and will be detected as a resistance change of the magnetic biosensor.

Resistance of the MTJ sensor having pinned layer magnetization angle fixed at $\theta_P=\pi/2$ and other to be rotating with an angle $\theta_F$ is given by [Sharad 2013]:

$$\Delta R / \Delta R_{max} = \frac{1}{2}\sin(\theta_F) \quad (3)$$

Where $\Delta R = R - R_O$ and $\Delta R_{max} = R_P \times TMR$ with $R_0$ and $R_P$ are resistances of TMR sensor at orthogonal and parallel configurations respectively. Sensing performance of the sensor can now be evaluated by computing the relative MR change parameter with and without bead as per following [Li 2003]-

$$\text{Relative MR Change} = \left[\left(\Delta R / \Delta R_{max}\right)_{with} - \left(\Delta R / \Delta R_{max}\right)_{without}\right]$$

$$= \frac{1}{2}\left[\sin(\theta_F)_{with} - \sin(\theta_F)_{without}\right] \quad (4)$$

Variation of Relative MR Change against free layer thickness for two different values of $t_P$ is depicted in fig.3 (a). It can be noticed from the figure that the MR change is higher for smaller $t_P$ because the dipole field from pinned to free becomes comparable to that of bead to free. Also increase in free layer thickness strengthen the bead's stray field mostly along the easy axis (+z direction) thereby decreasing its magnetization fluctuation and hence its field sensitivity. On the other hand, thicker $t_P$ exerts a greater dipole field on the free layer which makes the bead's field contribution on the free layer negligible, leading to a lesser MR change as compared to former case. However, strong dipolar effect from free to bead causes sensitivity to increase sluggishly with free layer thickness, $t_F$. Hence magnetoresistive device necessitates its two constituent ferromagnetic layers to have thickness close to each other for sensor related applications [Ferreira 2003, Albon 2009, Lian 2012]. The tunneling barrier thickness $t_{OX}$ of the MTJ can be optimized for maximizing TMR value and minimizing the R-C delay and hence the sensor response time as illustrated in fig. 3 (b) [Sharad 2013].

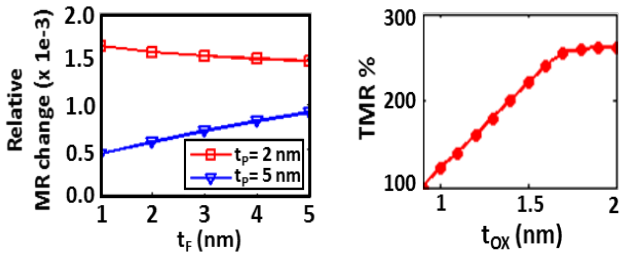

Fig. 3. (a) Relative MR Change with free layer thickness for 100 nm bead size using parameters as listed in table 1. (b) Variation of TMR VS. $t_{OX}$ for pinned layer parallel to free layer.

### B. SHM Based Assist

Basic structural arrangement of the proposed SHM based biosensing scheme has been shown in figure 4. Dimensions of the Biosensor were chosen to be in nanometer scale so that the spatial variation of magnetization within free layer can be neglected and mono domain micromagnetic modeling can be applied to it.

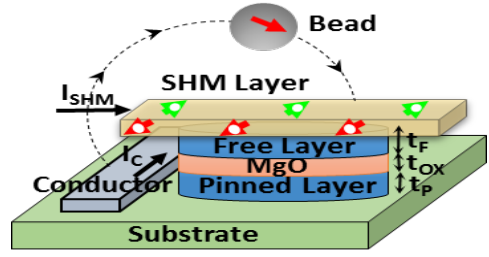

Fig. 4. Schematic drawing of the layered MTJ sensor structure with spin accumulation in SHM-magnet interface.

Dynamics of the free layer magnetization with normalized magnetic moment ($\hat{m}_F = M_F / M_{SF}$) with respect to saturation magnetization $M_{SF}$ can be described by the stochastic Landau-Lifshitz-Gilbert (LLG) equation [Behin-Aein 2010]-

$$(1+\alpha^2)\frac{d\hat{m}_F}{dt} = -|\gamma|\left[\hat{m}_F \times (\vec{H}_{eff} - \frac{\alpha}{|\gamma|}\vec{\tau}_{SHE}) + \alpha\hat{m}_F \times (\hat{m}_F \times \vec{H}_{eff})\right] + \vec{\tau}_{SHE} \quad (5)$$

Where, $\vec{H}_{eff}$ is the effective magnetic field on the free layer obtained by solving equation 2. Spin orbit torque $\vec{\tau}_{SHE}$ acting on the sensor magnetization due to the charge current density $J_{SHM}$ passing through SHM-nanomagnet interface can be expressed as [Khvalkovskiy 2013]-

$$\vec{\tau}_{SHE} \cong -|\gamma|\frac{\hbar J_{SO}}{2M_{SF}t_F}(\hat{m}_F \times \hat{m}_P \times \hat{m}_F) \quad (6)$$

Solution of the LLG equation provides time evolution of $\theta_F$ and its maximum fluctuation before reaching the stable state can be considered as the desired parameter for our calculations.

### C. Stress Based Assist

In this scheme, external voltage induced strain mediated magnetization fluctuations in the magnetostrictive layer has been exploited to improve the sensitivity of a magnetic biosensor. The structural arrangement of the proposed strain based biosensor is shown in figure 5.

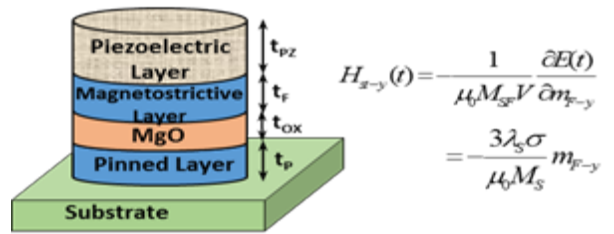

Fig. 5. A layered structure of the strain mediated sensor device with the stress induced magnetic field term.

Under this configuration, time evolution of the free layer dynamics has been captured by solving the LLG equation incorporating the stress induced magnetic field $\vec{H}_{st-y}$ [Fashami 2011] to the $\vec{H}_{eff}$ term in equation 5 while removing the effect of $\vec{\tau}_{SHE}$.

## III. RESULTS AND DISCUSSION

Analytical simulations were carried out to investigate the influence of spin hall effect based hard axis assist on improving the sensing performance of single MNP based MTJ Biosensor. All the parameter values used for single domain magnetic simulations are listed in the table 1 unless mentioned otherwise.

Table 1. Parameters for Magnet and LLG equation.

The effect of spin hall current on the sensor response has been investigated in fig. 6. Interfacial spin accumulation on the SHM exerts a Spin Orbit (SO) torque on the magnetic free layer along the +y direction. As a result of that magnetization $M_F$ is forced to move

| symbol | Quantity | Value |
| --- | --- | --- |
| $t_F$ | Free layer thickness | 1 nm |
| $t_P$ | Pinned layer thickness | 2 nm |
| $t_{OX}$ | Spacer layer thickness | 1 nm |
| $l \times w$ | Nanomagnet area | $10^4$ nm$^2$ |
| $\alpha$ | Gilbert damping coefficient | 0.05 |
| $\gamma$ | Gyromagnetic ratio | 2.21 x 10$^5$ rad.s$^{-1}$.T$^{-1}$ |
| $\Phi_H$ | Spin Hall angle | 0.24 |
| $M_{SF}$ | Free layer saturation Magnetization | 8 x 10$^5$ A/m |
| $M_{SP}$ | Pinned layer saturation Magnetization | 15 x 10$^5$ A/m |

away from easy axis, thus making a larger value of $\theta_F$. Figure reveals that 800 µA of $I_{SHM}$ can achieve ~ 6.5 times improvement in sensitivity for 100 nm bead at a distance of 500 nm from the sensing layer. It is also evident from the figure that stronger the bead interaction with free layer (by reducing height, h), more the $M_F$ will become unstable and hence will feel the effect of spin hall induced torque $\vec{\tau}_{SHE}$ greater towards its sensitive direction.

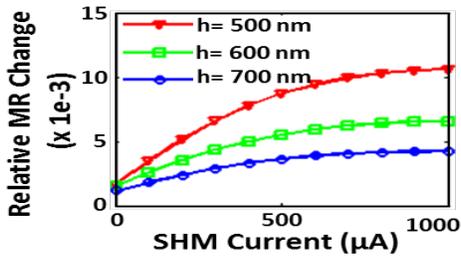

Fig. 6. Variation of Relative MR Change with SHM current for different bead height.

Increase in Nanomagnet area makes pinned layer magnetic moment stronger which in turn inhibits the free layer form sensing the magnetic bead. Hence smaller sensor dimensions are preferable for efficient detection. Figure 7 shows that larger bead size at a particular height, h exerts stronger field on the free layer and greater sensitivity can be observed. However, the increase in bead height from the sensor surface weakens bead's dipolar coupling on free layer thereby reducing the relative MR change.

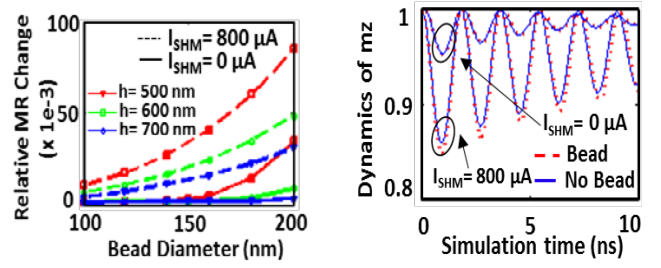

Fig. 7. Gain plot of the proposed mixer circuit for different values of PMOS width.

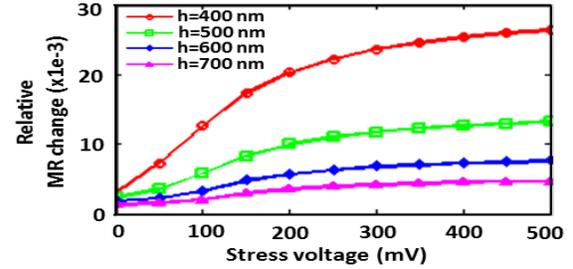

Fig. 8. Variation of Sensitivity with stress voltage for different bead height.

In the strain based design, a mono domain simulation has been performed to capture the effect of uniaxial stress on the stray field sensitivity of the proposed CoFeB-PZT multiferroic stack based MTJ sensor using the parameters reported earlier [Wang 2005, Roy 2011]. The resultant uniaxial stress subsequently exerts an effective torque on the CoFeB layer of the MTJ along the in-plane hard axis direction causing fluctuation of the $M_F$ orientation. Figure 8 shows that a 6 time improvement in the sensitivity can be attained by applying a stress voltage of 0.5 V when the bead is at a height of 500 nm. However, a higher stress voltage will not aid to enhance the sensitivity further because the $M_F$ gets stabilized in the y direction.

## IV. CONCLUSION

In conclusion, the concept of a simple and efficient MTJ biosensing scheme has been proposed. We analyzed the influence of SHE induced spin current and strain mediated effective field on the rotation of the free layer magnetization $\mathbf{M_F}$ towards the in-plane hard axis individually. Combination of high TMR and hard axis assist leads to a significant improvement in the output signal amplitude which may avoid the requirement of further circuitry for signal amplification thereby reducing the area and power consumption at circuit level while providing better sensitivity.